\documentclass[prl,amssymb,twocolumn,tightenlines,showpacs]{revtex4}

\usepackage{graphicx}
\usepackage{amsmath}
\begin{document}

\title{Nonlinear Interferometry via Fock State Projection}

\author{G. Khoury}
\email{khoury@physics.ucsb.edu}
\affiliation{Department of Physics, University of California,
Santa Barbara, California 93106, USA}

\author{H.S. Eisenberg}
\altaffiliation{Racah Institute of Physics, Hebrew University of Jerusalem, Jerusalem 91904, Israel}
\affiliation{Department of Physics, University of California,
Santa Barbara, California 93106, USA}

\author{E.J.S. Fonseca}
\altaffiliation{Departamento de F\'isica, Universidade Federal de Alagoas, Cidade Universit\'aria, 57072-970, Macei\'o, Alagoas, Brazil}
\affiliation{Department of Physics, University of California,
Santa Barbara, California 93106, USA}

\author{D. Bouwmeester}
\affiliation{Department of Physics, University of California,
Santa Barbara, California 93106, USA}


\begin{abstract}

We use a photon-number-resolving detector to monitor the photon-number distribution of the output of an interferometer, as a function of phase delay. As inputs we use coherent states with mean photon number up to seven. The postselection of a specific Fock (photon-number) state effectively induces high-order optical nonlinearities. Following a scheme by Bentley and Boyd [S.J. Bentley and R.W. Boyd, Optics Express \textbf{12}, 5735 (2004)] we explore this effect to demonstrate interference patterns a factor of five smaller than the Rayleigh limit.

\end{abstract}

\maketitle

Classical optics studies light fields containing macroscopic numbers of photons, while quantum optics mainly focuses on fields in which the mean photon number is much less than one. Light fields in between these two extremes have been relatively unexamined, because most detectors that are sensitive to single photons cannot distinguish one photon from two or more. Photon number resolving detectors have been developed, and have recently been applied to studies in quantum optics. In particular the Poisson photon number distribution of a coherent state has been confirmed \cite{TES, Lincoln} and non-classical photon statistics from parametric down conversion have been observed \cite{Waks}. It should be mentioned that it is also possible to reconstruct the photon number distribution of a light field without photon number resolving detectors, either with quantum homodyne tomography \cite{Smithey, Zavatta}, or with avalanche photodiodes of varying efficiency \cite{Paris, Paris2}.

In previous experiments, whether performed with single or multi-photon detectors, static field distributions have been measured. Here we report on the use of multi-photon detectors to monitor, in real time, the output of a Mach-Zehnder interferometer as we scan its phase. The inputs to the interferometer are coherent states with mean photon number up to seven. Projecting the output onto Fock states reveals a myriad of highly nonlinear responses hidden behind the familiar, linear, measurement of the mean photon number. This is a canonical example of measurement induced optical nonlinearities, which have already found applications in quantum information processing with linear optics \cite{Knill, Kok} and in a demonstration of a quantum controlled-NOT gate \cite{OBrien}. We exploit these hidden nonlinearities to simulate an N-photon absorber, a substance in development \cite{Wu, Witzgall, Cumpston} that has gained interest for its application in quantum photolithography \cite{Boto}. In this manner, we implement the scheme of Bentley and Boyd \cite{Bentley} to obtain interference patterns at one fifth the Rayleigh limit with classical light.

Our detector is the visible light photon counter (VLPC) \cite{Petroff, Turner, Lincoln}, which was developed for a particle tracking system at Fermilab based on photon generation in scintillating fibers. While similar in its operation to an avalanche photodiode, it differs in two important respects. First, it is sensitive to photons with wavelengths up to ${\sim}$\,30\,$\mu$m, so that it must cooled to 7\,K and extensively shielded from the large background of room temperature photons. Second, due to the details of the VLPC design, an incident photon generates a tightly localized avalanche of roughly 40,000 electrons. Since the active area of the device is much larger than the size of an avalanche, photons that are absorbed in different locations within the detector generate independent avalanches. One can determine the number of absorbed photons in a light pulse by measuring the height of the current pulse from the detector.

\begin{figure}
  \includegraphics[width=86mm]{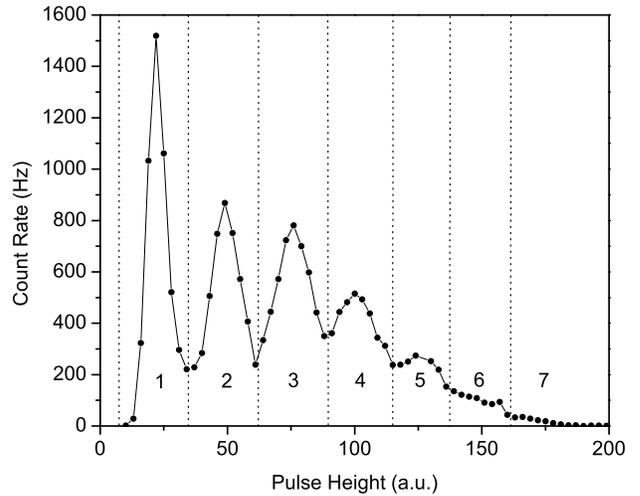}\\
  \caption{Example pulse height distribution of the VLPC. Vertical dashed lines show decision thresholds. The single photon count includes dark counts.}\label{PHD}
\end{figure}

Figure \ref{PHD} shows a representative pulse height histogram from the VLPC. The peaks from absorption of one, two, \textit{etc}. photons can be distinguished. Their nonzero width is due predominantly to fluctuations in the number of electrons produced per photon. This multiplication noise makes the peak for absorption of $n$ photons $\sqrt{n}$ broader than the single photon peak, because it is the result of $n$ independent processes.

In the multi-photon resolving experiments mentioned above, the photon number distribution was determined by sending all current pulses to a multichannel analyzer that measured the pulse height histogram. The photon number distribution was then determined by fitting the histogram to a sum of Gaussians. This is successful, but the use of a multichannel analyzer limits the allowable data rate. Also, the determination of how many photons were absorbed is made only in the fitting stage, after the experiment is over. We employed a different method that allowed us to determine, on a shot by shot basis, how many photons were detected. We used several comparators (Analog Devices AD96687) that compare the amplified VLPC current pulse to seven variable threshold levels $L_k$. We first measured a pulse height histogram by scanning two closely spaced levels and monitoring the rate of pulses between them. We then set the thresholds of the comparators to the minima between the photon peaks, as shown by the vertical lines in Fig. \ref{PHD}. A pulse that lay between levels $L_k$ and $L_{k+1}$ was recorded as the detection of $k$ photons. Due to fluctuations in the number of electrons per avalanche, it is possible that $k$ photons can be recorded as $k + 1$ or $k-1$. This becomes more of a problem at higher $k$, as the peaks overlap more. In the experiments reported here this overlap is reasonably small for $k<5$. If more accuracy is required, this can be corrected if one has all the nonzero count rates \cite{WaksIEEE}.

\begin{figure}
  \includegraphics[width=86mm]{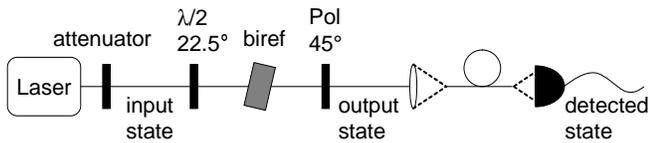}\\
  \caption[Experimental Setup]{Experimental Setup. The waveplate, birefringent crystal (biref), and polarizer form a polarization Mach-Zehnder interferometer. The output state is collected into a single mode fiber and sent to the VLPC.}\label{MZdiagram}
\end{figure}

Our optical setup is shown in Figure \ref{MZdiagram}. Our light source was a 780\,nm wavelength, 100\,fs pulsed Ti:Sapphire laser. Because our detector begins to saturate at around 5\,MHz, we reduced the repetition rate $R_\text{rep}$ of the laser to 40\,kHz with a Q-switched amplifier system. The laser light was heavily attenuated with a neutral density filter and sent into a polarization Mach-Zehnder interferometer. The phase delay of the interferometer was changed by tilting a birefringent crystal inside it. The light from one output was then collected into a single mode fiber and sent to the VLPC.

\begin{figure}
  \includegraphics[width=86mm]{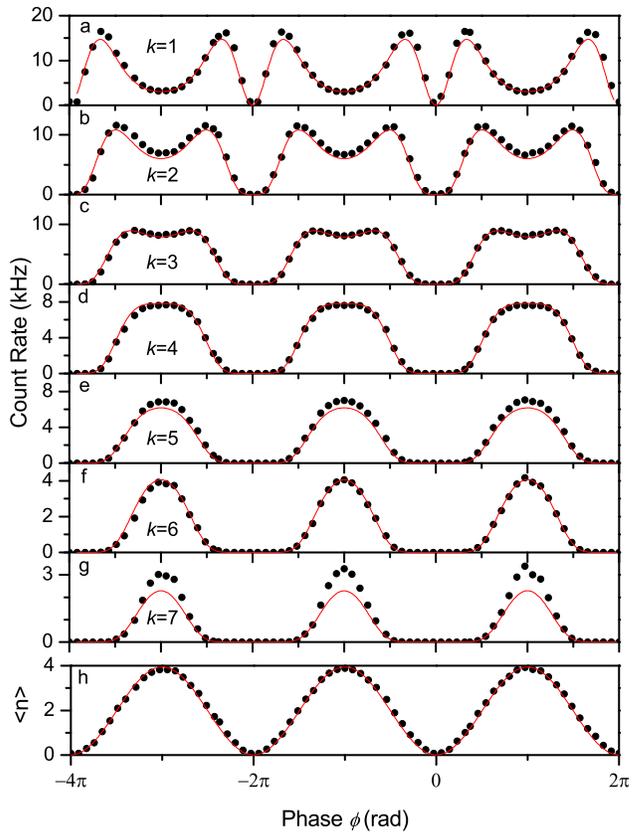}\\
  \caption{(Color online)(a) through (g): Count rates \textit{vs.} phase
           for the detection of one to seven photons
           from an input state with $n_0=3.95$. (h): average photon number computed from
           weighted sum of count rates. Solid lines show theoretically
           expected values. }\label{WFALL}
\end{figure}

For our purposes \cite{Molmer}, we can take the laser light to be in a coherent state $| \alpha \rangle$, given by \cite{Loudon}
\begin{equation}
  | \alpha \rangle = \sum_{k=0}^\infty e^{-|\alpha|^2 / 2}
  \frac{\alpha^k}{\sqrt{k!}} | k \rangle ,
\end{equation}
where $| k \rangle$ is a $k$-photon Fock state and $\alpha$ is the dimensionless electric field amplitude of the coherent state. The mean photon number of such a state is $\langle n \rangle = |\alpha|^2$. The interferometer transforms the input state $| \alpha \rangle$ into the output state $| \alpha \sin{(\phi/2)}\rangle$, so that the intensity of the output varies sinusoidally with phase. Losses in the subsequent optics and detector transform the interferometer output into the detected state $| \eta \cdot \alpha \sin{(\phi/2)}\rangle$, where $\eta^2$ is the total detection efficiency of the setup. Thus the detected state is equivalent to the output of an interferometer fed by a weaker input state.

Figures \ref{WFALL}(a) to (g) show the count rates for a maximum average detected photon number $n_\text{max}$ of 3.95. Each plot shows the count rate for detection of $k$ photons, or equivalently the overlap of the detected coherent state with the $k^\text{th}$ Fock state. An important feature of each count rate is the non-sinusoidal behavior. For photon numbers $k$ less than $n_\text{max}$ there is a local minimum at the position of maximum light intensity. Ideally, the $k$-photon count rate $R_k$ is given by
\begin{eqnarray}
  R_k &=& R_\text{rep} | \langle k | \alpha \sin (\phi/2) \rangle |^2  \label{RK} \nonumber \\
      &=& R_\text{rep} e^{-n_\text{max} \sin^2 (\phi/2)} \frac{( n_\text{max} \sin^2 (\phi/2) )^k}{k!}.
\end{eqnarray}
As the intensity of the light increases, its overlap with the $k$-photon state increases only until $\langle n \rangle = k$. Because $R_k$ is maximal when $\langle n \rangle = k$, the number of count rates that peak before $\phi = \pi$ is an indication of $n_\text{max}$. To determine this value quantitatively, we fit the single photon count rate to equation \ref{RK}, which yielded $n_\text{max}=3.95$. Using this value we generated the lines for $R_2$ to $R_7$ and $\langle n \rangle$. Note that $R_7$ is significantly higher than predicted from equation \ref{RK}, because eight photon absorption is non-negligible.

As can be seen from their complicated phase dependence, the individual count rates are composed of many frequency harmonics. However, when they are combined as $\sum_{k=0}k p_k$ as in Figure \ref{WFALL}h, only the fundamental frequency remains. It is this sinusoidal sum that is usually associated with the output of an interferometer. By monitoring the variation of the photon number distribution, the rich underlying behavior is brought to light.

\begin{figure}
  \includegraphics[width=86mm]{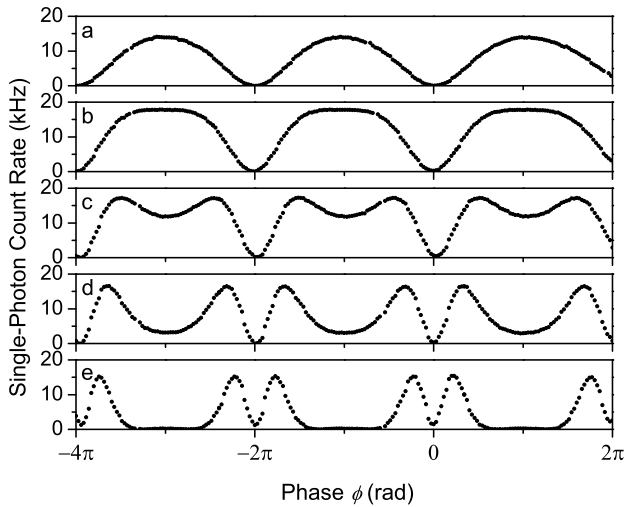}\\
  \caption{Single photon count rate \textit{vs.} interferometer phase at
  several mean photon numbers: (a) 0.463, (b) 0.99, (c) 1.99, (d) 3.95,
  (e) 7.18.}\label{onesvspower}
\end{figure}

To observe the appearance of a dip in the single photon count rate, we took five scans with different values of $n_\text{max}$. Figure \ref{onesvspower} shows the observed single photon count rates. At low photon number, the single photon rate follows the sinusoidal intensity pattern, but as $n_\text{max}$ increases, it smoothly deforms into a double peak structure.

Given that $R_7$, for example, is much more sharply peaked than $\langle n \rangle$, and therefore contains many higher frequency harmonics, one might expect that an interferometer can be made more sensitive by monitoring $R_7$ rather than $\langle n \rangle$. This is not the case. To prove this, we need a result from quantum estimation theory \cite{Dowling}. When measuring a quantity $\hat{O}$ to estimate a parameter $\phi$, the uncertainty in the estimate is quantified by
\begin{equation}
  \Delta \phi = \frac{\Delta \hat{O}} {\big|\frac{d\langle \hat{O} \rangle}{d\phi}\big|},
\end{equation}
where $\Delta$ is the standard deviation in the measurement of a quantity. The standard method of interferometry is to measure $\hat{n}$. For a coherent state with mean photon number $n_\text{max}$,
$\Delta \phi = 1/{\sqrt{n_\text{max}} \cos (\phi/2)}$.
The best sensitivity is obtained at $\phi = 0$, where the intensity is zero.

The analysis for measuring a $k$-photon count rate is similar. The number of counts we get is given by a binomial distribution, and the variance in the count rate is $(\Delta R_k)^2 = p_k(1-p_k) / N$, where $N$ is the number of laser pulses used, and $p_k$ is the probability that the state projects to $|k\rangle$. This leads to
\begin{equation}
\Delta \phi = \frac{2\sqrt{1/p_k - 1}}{\sqrt{N} \left( \frac{k}{n_\text{max} \sin^2 (\phi/2)} -1 \right) n_\text{max} \sin (\phi)}.
\end{equation}
The phase setting that gives optimum sensitivity is now a function of $n_\text{max}$ and $k$. Sensibly, the worst sensitivity occurs when $\langle n \rangle = n_\text{max} \sin^2 (\phi/2) = k$, where $R_k$ is maximal and thus insensitive to changes in the output intensity.

\begin{figure}
  \includegraphics[width=86mm]{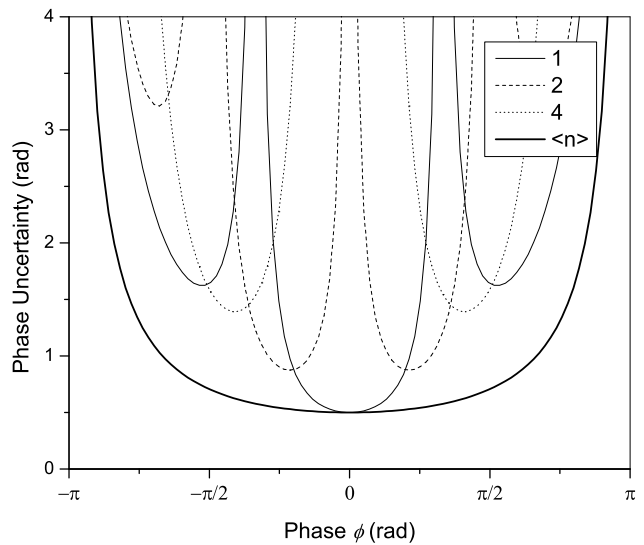}
  \caption{Calculated phase sensitivity \textit{vs.} phase when measuring the average
  photon number, and the overlap of the state with the one, two, or four
  photon Fock states for $n_\text{max}=4$.}\label{focksens}
\end{figure}

Figure \ref{focksens} compares the calculated sensitivity of these two measurement schemes. The usual intensity measurement is significantly more sensitive than monitoring a $k$-photon count rate, except for the single photon count rate, which yields equal sensitivity at $\phi=0$. Here the output intensity is nearly zero, there are never any pulses with two or more photons, and the single photon rate is proportional to the intensity.

\begin{figure}
  \includegraphics[width=86mm]{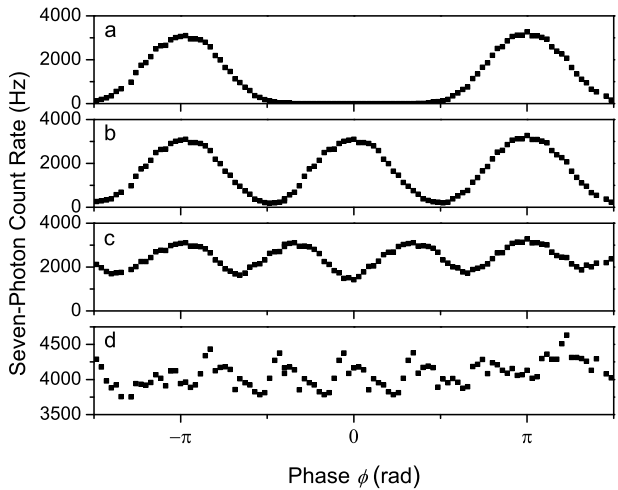}\\
  \caption{Seven photon count rate \textit{vs.} interferometer phase,
  (a) alone, and superimposed with copies shifted by (b) $\pi$, (c) $2\pi/3$, and (d)
  $2\pi/5$. Note the change in scale in part d.}\label{byn}
\end{figure}

The presence of these higher harmonics is the result of the nonlinearity of the detection process, \textit{i.e.} the multi-photon absorption probability is a highly nonlinear function of the light intensity. Materials that absorb nonlinearly could prove useful for many tasks such as photolithography. A recent proposal by Boto \textit{et. al.} \cite{Boto} suggests a way to obtain interference features a factor of $N$ smaller than the Rayleigh limit. It requires both an $N$-photon absorbing resist and light in the entangled state $| N \rangle | 0 \rangle + e^{iN\phi} | 0 \rangle | N \rangle$, where the kets denote the two interfering light modes, and $\phi = 2 \pi x / \lambda$ is the phase difference between them at a position $x$ on the resist. In this case the probability of absorbing $N$ photons goes as $\cos^2(N\phi/2)$, which yields interference fringes $N$ times finer than the Rayleigh limit. Later, Bentley and Boyd \cite{Bentley} demonstrated that entanglement is not essential. Their scheme obtains interference fringes a factor of $N$ smaller than the Rayleigh limit using an $M$ photon absorber and classical light. The major drawback is that for high visibility fringes, $M$ must be much greater than $N$. The scheme was presented as follows: interfere two laser beams to produce a standard interference pattern on the photoresist, and increase the relative phase of the two beams by $2\pi/N$ for each pulse. The $M$-photon absorption is proportional to $\sum_{i=1}^N (E_i E_i^*)^M$, where $E_i$ is the electric field on the resist from pulse $i$. Shifting the phase of the pulses averages out the spatial frequencies on the resist below the desired frequency of $\lambda / N$.

Here we present a similar but experimentally easier to realize scheme. Figure \ref{byn}(a) shows the 7-photon absorption \textit{vs.} phase at $n_\text{max} = 3.95$. Panels (b), (c), and (d) show the results of a data post-processing in which the same pattern is superimposed with copies shifted by multiples of $2\pi/2$, $2\pi/3$, and $2\pi/5$, respectively. The essence of the proposal now becomes clear: the multiphoton absorption peaks on the resist are narrow, and more than one of them will fit in the space of one wavelength. The variation in the count rate that remains under shifting by $2 \pi / N$ is a measure of the harmonics at and above $N$.

In conclusion, we have presented real-time measurements of a photon number distribution using a detector that can resolve photon numbers up to seven. We used this technique to measure the output of an interferometer, which showed that the photon count rates behave highly nonlinearly and vary on a sub-wavelength scale. While the phase sensitivity of an interferometer cannot be improved by this method, we use it to demonstrate sub-wavelength interference patterns using only classical states of light and the inherent nonlinearity of detection.

We thank Don Lincoln and Alan Bross for loaning us the VLPC's. This research has been supported by NSF Grant No. 0304678 and by the DARPA MDA 972-01-1-0027 grant. EJSF is grateful for support from Instituto do  Mil\^enio de Informa\c{c}\~ao Qu\^antica, CNPq, FAPEAL, CAPES. H.S.E. acknowledges support from the Hebrew University.

\end{document}